\documentstyle[11pt]{article}
\begin{document}
\vbox{
\begin{flushright}
AZPH-TH/94-15
\end{flushright}
\title{Second-Order Scaling in the Two-Flavor QCD Chiral Transition}
\author{Arjun Berera* \\
Department of Physics \\
University of Arizona \\
Tucson, Arizona 85721}
\date{}
\maketitle
\begin{abstract}
Scaling behavior is analyzed for the two-flavor QCD lattice gauge theory
chiral transition.
Leading scaling behavior
and correction to leading scaling from lattice spacing effects are
examined for the quark condensate.
Scaling predictions under the assumption of
quark mass dominance
are tested for the longitudinal correlation
length.  Second order scaling behavior is consistent with present
data.
\end{abstract}
\vspace{20mm}}


\bigskip
\eject

\noindent {\Large\bf I. Introduction}
\medskip

Some time back it was pointed out by Pisarski and
Wilczek \cite{pisarski},
that the QCD chiral phase transition has
a suitable order parameter from which an effective,
Landau-Ginzburg-Wilson, critical theory could be constructed.
Recently Wilczek \cite{wilczek} has examined the case of two fermion flavors
in the chiral limit,
where the effective LGW-theory is now the
SU(2)$\times$SU(2)=O(4) Heisenberg model.  This latter model
has been well studied for spin systems in statistical
mechanics.  As such,  the necessary information regarding
the critical behavior of this theory was immediately available.
Subsequently Rajagopal and Wilczek \cite {rajagopal}
gave a transcription dictionary
which related the O(4)-Heisenberg transition into
the language of lattice gauge theory.  In particular
the quark mass and quark condensate transcribed to the magnetic
field and magnetization respectively in the language of spin
systems.
This transcription now readily allows testing the original
assertion, that the chiral phase transition is
second order, by checking for universal scaling behavior
and comparing it to that of the O(4)-Heisenberg model.

As lattice simulation data on the SU(2)-chiral
transition increases, it will become
possible to analyze the scaling behavior near the tentative transition point.
Such tests will provide important input in establishing the correctness
of the above hypothesis by \cite{wilczek}
that the transition is second order and in
the O(4)-Heisenberg universality class.  To obtain reliable estimates from such
analyses, a good understanding is first needed of the relevant scales involved
in real finite temperature simulation studies.
In general there are four scales involved here, which in descending order
are the length of the lattice box L, the correlation length $\xi$,
the QCD scale $\Lambda_{QCD}$, and the lattice spacing a.
Ideally, for studying critical phenomena scaling,
one would like $L>>\xi>>\Lambda_{QCD}>>a$.
In real simulation, typical hadron sizes span only a few lattice
spacings, and only two or three hadrons fit inside the lattice volume.
This situation places greater problems in studies of critical phenomena,
above those well familiar from zero temperature studies.
Firstly the divergence of the correlation length is impeded by
the lattice volume, leading to finite size effects.  In addition,
since asymptotic scaling in the ultraviolet is only approximate,
hadron states exhibit a-dependence,
which in turn complicates infrared critical
phenomena scaling through the addition
of correction terms
\cite{berera}.
Furthermore, the two problems are complementary in a fixed
lattice volume.  For, to improve
asymptotic scaling by decreasing the physical size of a lattice spacing,
(ie placing more lattice sites inside a hadron state), goes at
the expense of decreasing the maximum physical size that the correlation
length can achieve.

To overcome such difficulties, there is some consideration \cite {karsch}
that it is more advantages to examine critical
phenomena scaling at finite, not vanishing, quark mass.
In such a case, for a sufficiently large quark mass,
scaling theory tells us that the correlation length behaves
as$^{**}$
$$
\xi = {A \over {h^{\nu/\Delta}}} (1 + a_1(t/h^{1/\Delta})+ ...)
\ , \eqno{(1.1)}
$$
so is cutoff by a scale set by the quark mass.  Here $\Delta$ is the gap
exponent, related by the scaling laws to $\delta $ and $\beta $
by $\Delta=\beta \delta$.  For the O(4) system $\beta=0.38 \pm 0.01$,
$\delta = 4.82 \pm 0.05$ \cite{baker} so that
$\Delta =1.83 \pm 0.06$.

However, the more one suppresses the correlation length, the more significant
are the effects from the microscopic scales.   The dominant contributions
are from dimension four operator corrections arising from the microscopic
details of the system.  In particular, these corrections contain the dominant
effect of the nonzero lattice spacing.

Let us estimate the size of this effect on the magnetization.
The scaling form for the magnetization is \cite {berera2}
$$
M=Bh^{1/\delta} f(|t|/h^{\Delta})(1+b_1|t|^{\omega \nu}+...)
\ , \eqno {(1.2)}
$$
and at t=0 it is
$$
M=Ch^{1/\delta}(1+c_1h^{\omega \nu \over \Delta}+ ...)
\ . \eqno{(1.3)}
$$
The second term inside the parenthesis
in both equations above, is the correction
to leading scaling term arising from a nonzero lattice spacing.
Here, for the O(4) system, $\nu =0.73 \pm 0.02$ \cite{baker}
and $\omega =0.46$ \cite{brezin1,ma,bervillier}.
Away from $t=0$, when $t/h^{\nu/\Delta}>>1$ recall $\xi \sim t^{-\nu}$.
At $t=0$, the magnetic field will dominate, where from
Eq. (1.1) we find $\xi \sim 1/h^{\nu/\Delta}$.  In both cases the correction
to scaling term goes
as $\xi^{-\omega}$.  Let us suppose that the correlation length
is ten lattice units long, a size
much bigger than found in simulations.  If we assume that the coefficients
C and $C*c_1$ for the leading and next-to-leading scaling terms respectively
in Eq. (1.3) are of about the same magnitude, then the latter
term gives about a 30 percent correction.

In this paper we will study scaling behavior.  Due to limited
data from lattice gauge simulations, we can not call this a
legitimate scaling analysis.  A legitimate scaling analysis would
require demonstrating a consistent fit of many data points
to expected scaling forms.  At present we are working with a hopeless
three or four points in any single type of scaling analysis.
However, there are a few qualitative and also quantitative
predictions from scaling theory that can be tested
with sparse data.  One such qualitative prediction
is  the correction to leading scaling from effects of a nonzero
lattice spacing.  Lattice spacing effects have frequently been
cited in various studies \cite{gottlieb2,engels,kapusta}.
The presence of a nonzero lattice spacing necessarily implies
that there will be corrections to leading scaling.  The theory of
second order critical phenomena predicts well prescribed
functional forms for these correction terms.
In section 2, under the assumption that the chiral transition
is second-order, we will examine those corrections to leading scaling
behavior in the magnetization, which
arise from a nonzero lattice spacing.
Quantitatively, scaling behavior of the
longitudinal correlation length is a
key indicator and can be tested with present data.
In section 3 we will make estimates of its scaling behavior
based on lattice simulation data.

\medskip
\noindent {\Large\bf II. Magnetization}
\medskip

Lattice simulators have been examining the magnetization,
$<\bar \psi \psi>$, for some time now.  There appears sufficient
consistency in their results to warrant closer analysis.
In figure (1) the plotted points are from simulation data in
\cite{gottlieb,mawhinney,hemcgc}.
The solid line is a fit to Eq. (1.3) without the leading scaling correction
term.  The dashed curve is a fit with the correction term added.
Both are for the O(4) exponents.  A similar fit to O(2) exponents
gives minute differences, undetectable on the graph.

The results we present for the magnetization demonstrate
the effect of the
lattice spacing in the scaling forms, when everything else is
kept the same.
Rather than extrapolating a zero magnetic field critical temperature,
we allowed for the analytic shifts that arise in the presence of
a magnetic field.  The exact value of these transition
points would require outside nonperturbative information
which is not calculable at present.
In our fitting procedure, we simply estimated the transition point
at each $m_q$, as the place
of maximum drop-off in the magnetization profile as a function
of $\beta$. Let us justify
the sufficiency of this crude method for present purposes.
In a more rigorous treatment one may make a best fit with the critical
temperature as an initial unknown fitting parameter.  A suggestive
form for such a fit would be the leading expansion to Eq. (1.2),
$$
M=Ah^{(1/\delta)}(1+a_1 |t|/h^{1/\Delta})
\ . \eqno{(2.1)}
$$
By this approach,
one would be committed to a single critical
temperature for all values of $m_q$.  One may worry about
analytic corrections
to $\beta_c$ as a function of $m_q$.
This concern may be further extenuated
here since our "temperature" parameter.
$\beta = 6/ g^2$ is itself analytically related to the physical
temperature parameter.  From present data, indications are that the variation
of $\beta_c$ is only about 2 percent for quark masses
ranging from 0.004 to 0.0125.
To be precise, $\beta_c$=5.48, 5.475, and 5.54 for
$m_q$=0.004, 0.00625 and 0.0125 respectively.  The outer two are
given in \cite{mawhinney} and \cite{gottlieb} respectively and
the middle point is from our estimate.
At a later stage, when more data points are available, it
may make sense
to use the form (2.1) at each $m_q$ so as to also
make $\beta_c(m_q)$ a fitting parameter.

One point which becomes clear from the above examination
of the magnetization is
that it will be difficult to differentiate between O(2) and O(4)
critical behavior even with increased amount of data.
This is because lattice correction effects dominate.
A tractable question that one can hopefully decide upon
is whether the transition is second order.  Here general properties
of scaling forms, if consistently matched with data, will be a significant
advance.  In particular one can pose the question in a simplified
fashion for present needs as deciding whether the transition
is first or second order or whether it is some
sort of smooth transition.  Application of scaling theory
should be sufficient to decide between the first
two options.  Establishing that the longitudinal
correlation length diverges should place considerable doubts on the
latter possibility.

\bigskip
\noindent {\Large\bf  III. Longitudinal Correlation Length}
\medskip

Let us now turn to the longitudinal correlation length, which in
the lattice gauge theory language is the inverse sigma screening mass,
$1/m_\sigma$.  There are concerns \cite{doug}
that for $m_\sigma$ only the isospin I=1
channel contributed in the lattice gauge measurements of
\cite{gottlieb,bernard}.
This, in particular, means that the ground state I=0 sector is not
contributing in the measurement of $m_\sigma$.  For zero temperature
properties in the scalar sector, this elimination of the I=0
sector would be a cause for obvious concern.  Near the critical
point, one expects scalar particles from all low lying
isospin sectors to contribute
with equal importance.  Under this assumption we expect the universal
scaling properties
of critical behavior to still be reflected in $m_\sigma$.
We follow this reasoning and assume in the analysis below that ${1 \over
{m_\sigma}}$, as measured by lattice simulations in \cite{gottlieb,bernard},
is the longitudinal correlation length.

A qualitative assessment of the longitudinal
correlation length in \cite{gottlieb} shows that it
increases a little as one approaches $\beta_c$ from below.
In the case of the hot start, it peaks near
$\beta_c$.  It then drops a bit as one goes to higher $\beta$
and finally levels off.  In the cold start case, although there
is no peaking, there is still a definite rise as it approaches $\beta_c$
and then a leveling of.

In a system nearing a second order phase transition,
if the diverging correlation length is influenced by the box size, it
is seen by its sharp rise being cutoff
at some intermediate value.
In this region
the finite extent of the box modifies its behavior
to \cite{brezin4},
$$
\xi=BL(1+b_1|t|L^{1/\nu}+ ...)
\ . \eqno(3.1)
$$
If in the same system, a large enough magnetic field is imposed,
the correlation length is also flat near $\beta_c$, as seen from Eq. (1.1).
Also for comparison, if the temperature
dominates over the volume and quark mass,
the simplist form consistent with
scaling theory is,
$$
\xi = a |t|^{-\nu}
\ . \eqno(3.2)
$$
Using the data in \cite{gottlieb}, which is fairly representative
of lattice simulation data for $m_\sigma$,
the divergence of the longitudinal correlation length evidently
is cutoff as it approaches $\beta_c$.  We anticipate
that either the volume or quark mass controls the correlation
length near $\beta_c$.  We will define a region as quark mass
(volume) dominated, when the divergence of the correlation length
as $t \rightarrow 0$, is stopped due to the presence of a quark
mass (a finite volume).  Thus in the quark mass dominated region,
for example, if one lowers the quark mass, it will increase $\xi$
as $t \rightarrow 0$.  However, if one increases the volume and keeps
the quark mass fixed, $\xi$ will not increase as $t \rightarrow 0$.

To judge between regions of
volume and quark mass dominance,
we first observe from Eq. (1.1) that
$$
{\xi(2m_q) \over \xi(m_q)} ={1 \over {2^{\nu/\Delta}}}=1/2^{0.4}=0.76
\ , \eqno(3.3)
$$
where the exponent used on the right hand side is for the O(4) system.
There is some data at two values of quark mass with fixed box
size L=12 and $N_t=6$ in \cite{bernard}.  Using the critical temperatures they
estimated from
the magnetization profiles, $\beta_c(m_q=0.0125)=5.42$
and $\beta_c(m_q=0.025)$=5.445, we find the ratio of the
measured correlation
lengths at the respective critical points to be,
$$
{\xi_m(0.025) \over \xi_m(0.125)}=
{{0.65 \pm 0.03} \over {0.839 \pm 0.009}} =0.77 \pm 0.05
\ , \eqno(3.4)
$$
which agrees with eqs (3.3).  Above the subscript m is to
indicate measured value.  The closeness of the two ratios in
Eqs. (3.3) and (3.4) should
not be taken too seriously considering the approximations involved.
Nevertheless, from a qualitative inspection of the data
in the transition region,
it is evident that the correlation
length at $m_q=0.0125$ is larger by roughly the above amount.
This suggests that down to $m_q=0.0125$
the quark mass effects dominate the finite size effects
at L=12.  This offers numerical support to one of the assumption
on which the recent analysis by Karsch \cite{karsch} is based.

We can also make a comparison of $\xi(\beta_c)$ at $m_q=0.0125$
for lattice sizes of L=12 and 16 from the data in \cite{bernard} and
\cite{gottlieb} respectively.
The former is for $N_t=6$ and the latter is for $N_t=8$.
{}From this comparison, one can assess the degree to which
the continuum limit
(i.e. ultraviolet scaling
limit, where the theory becomes independent of its short distance
cutoff)
has been reached near the transition
point.  To understand our reasoning
here, first recall that the critical point in both systems corresponds
to the same physical value for the temperature.
This implies that the physical scale
of a lattice unit in the two systems, as given by $a_{N_t}={1 \over
{N_t*T_c}}$,
is different.  Since our considerations below only involve
comparing two lattice systems, for convenience let us define
the lattice spacing in the $N_t=6$, L=12
system to be unity and refer to this as the physical unit of length.
Then in these units, if the continuum limit
has been reached, the $N_t=8$, L=16 system should  reproduce the same physics
when one sets the lattice spacing at 3/4.  In particular the system
size in lattice units of the $N_t=8$ system, $L_8$, is in physical
units
$$
L^{ph}_8={3 \over 4} L_8 \ . \eqno(3.5)
$$
For the two systems under comparison, this implies their
physical sizes are
the same.  Thus we can not study finite size scaling effects
in comparing these two systems.

Turning to the correlation length, define $\xi_8(t)$ $(\xi_6(t))$
as the correlation length in the $N_t=8$ ($N_t=6$) system in units
of its lattice spacing.
If the continuum limit has been reached, at zero quark mass
and at the same physical temperature, we expect to find
from measurement the relation,
$$
\xi_6(t)={3 \over 4} \xi_8(t) \ , \eqno(3.6)
$$
where t is the reduced temperature.
If, as in our case,
the same dimensionless magnetic field (quark mass) is applied to
the two systems, the above relation is modified.  One can intuitively
see this by noting that if the same dimensionless magnetic field
is applied on each lattice site, then the system with the greater
thickness (the $N_t=8$ system) feels a greater overall effect.
We can see this numerically by accounting for dimensions.
One has,
$$
m_q^{ph}={m_q \over a}
\ , \eqno(3.7)
$$
where a is the lattice spacing,
the superscript ph implies the quark mass in
physical units, and $m_q$ is the dimensionless quark mass.
Recalling that we are considering the lattice spacing in the $N_t=6$ system
as our
physical length unit, we imagine fitting the parameters in Eq.(1.1)
(ie A,$a_1$ etc...) in these units.  Let us now analyze
how well the quark mass is controlling the correlation length
in the two systems.  From Eq. (1.1) we expect,
$$
{\xi^{ph}_8(0) \over \xi^{ph}_6(0)}=({{m^{ph}_6} \over {m^{ph}_8}})^{\nu\over
\Delta}
=(3/4)^{0.399}= 0.89
\ . \eqno(3.8)
$$
{}From simulation data in \cite{gottlieb} we find at the critical point that
$^{***}$,
$$
\xi^{ph}_{8m}(0)={3 \over 4}\xi_{8m}(0)=(3/4){1 \over {0.561 \pm 0.005}}
\ , \eqno(3.9)
$$
where the subscript m is to indicate the measured value.
For the measured ratio at the critical point, it is,
$$
{\xi^{ph}_{8m}(0) \over {\xi^{ph}_{6m}(0)}} =
{({3 \over 4})}({{0.65 \pm 0.03} \over {0.561 \pm 0.005}})=
0.869 \pm 0.041
\ . \eqno(3.10)
$$
The agreement between eqs. (3.8) and (3.10) offers plausible support
for the joint assumption of quark mass dominance and continuum behavior.
This single comparison is insufficient to make any conclusions,
but it serves as an indicator based on the available data.

\bigskip
\noindent {\Large\bf IV. Conclusion}

The rise in the measured magnetization, as shown in figure 1, is too
sharp to fit leading scaling expectations.  We are able to account for
its behavior by a parametric fit to a scaling form that includes
effects of a nonzero lattice spacing.  Of course, to fit three data points
with two parameters is no feat and at present that is all we are able
to do.  Nevertheless, we still learn from this analysis that
leading scaling behavior is probably insufficient.  Also, by turning
to outside physics considerations as discussed
in \cite {berera} and finding support from our curve fit,
it tentatively appears that corrections to leading scaling from
a nonzero lattice spacing may be the explanation.

Turning to the issue of quark mass dominance,
in \cite{boyd} a relation was given that separates the regimes in which the
quark mass and the lattice size control the correlation
length.
{}From our findings, we can fit this equation to obtain
an upper limit on the
smallest
quark mass, which separates it from
the volume dominated region.
In lattice units of the $N_t=6$ system, we get
$$
m_q > CL^{-b} =6.1L^{-2.49}
\ , \eqno(4.1)
$$
where $b={{\gamma \delta} \over {\nu(\delta-1)}}=2.49$.

To the limited extent of our analysis, the longitudinal
correlation length apparently exhibits behavior characteristic to a second
order phase transition.  It should be monitored in future lattice gauge
simulations.  At present, the HEMCGC collaboration \cite{hemcgc}
is computing $m_\sigma$ at
$m_q=0.00625$ for a $16^3 \times 8$ lattice.  If we continue
to accept quark mass dominance and continuum behavior, similar to Eq.(3.3)
we expect from Eq. (1.1),
$$
{\xi(m_q=0.0125,t=0) \over \xi(m_q=0.00625,t=0)} =
{1 \over {2^{\nu/\Delta}}}=0.76
\ . \eqno(4.2)
$$
Should this fail,
the first explanation that we can offer
is a crossover from the quark mass dominated to the volume dominated
regime.  In the event of such a breakdown,
the result would still be useful in
obtaining a more definitive boundary in Eq. (4.1).
Before doing this, however,
should Eq. (4.2) not be confirmed by measurement,
larger lattices would have to be tried in order to be sure that
the cause is volume effects.  If we rescale Eq. (4.1) for the $N_t=8$
case, we expect that for $m_q=0.00625$ the mass dominated
region should occur for $L>19$, where this is in lattice units of an
$N_t=8$ system.  For lattices bigger then this, we minimally expect
Eq. (4.2) to hold.  If so, then one can reasonably assume that
simulation data is exhibiting the simplist form of second order
scaling.  This would be a marked advantage for lattice
theorists in further exploration of this region.
If the ontcome is not in line with the
above expectations, then the situation will need more
consideration, and the analysis of section 3 may be a fortuitous
coincident, but otherwise hollow without content.

\medskip
\noindent {\Large\bf Acknowledgements}

\medskip

\normalsize

I thank Professors Douglas Toussaint and Michael Wortis for helpful
discussions.  Some of the less documented facts about scaling, that
we used in the text, were taken from discussions with Professor
Wortis.
Financial support was provided
by
the U. S. Department
of Energy,
Division of High Energy and Nuclear Physics.

\medskip
*  Present address: Department of Physics, the Pennsylvania State University,
104 Davey Laboratory, University Park, PA 16802-6300

\medskip
** This relation can be obtained by standard scaling arguments
as follows.  We first write the asymptotic behavior of
the correlation length at zero field in the familiar form
$\xi \sim t^{-\nu}$ \cite{amit}.
In the presence of a finite magnetic field h,
although the nonanalytic behavior in t will remain the same,
$\xi$ in general can depend on h also.  We can write this as
$\xi \sim t^{-\nu} g'(t,h)$.  By the homogeneity hypothesis of
scaling, the dependence in fact can only be with respect to the ratio
${t \over {h^{\sigma}}}$.  This implies the general
form $\xi \sim t^{-\nu} g(
{t \over {h^\sigma}})$, where g(x) is of degree $\gamma$.
The ratio $t \over {h^\sigma}$ must be the same as in the singular part
of the free energy.  For the free energy it is well known that
$\sigma= 1/\Delta$ \cite{amit}.

In the region t=0, $h \neq 0$, $\xi$ is finite with a nonanalyticity
at h=0.  In order to satisfy this requirement, we must have
$t^{-\nu}({t \over {h^{1/\Delta}}})^\gamma={1 \over h^{\sigma^\prime}}$.
This relation implies $\gamma = \nu$ so that
$\sigma^\prime={\nu \over \Delta}$.  Extracting this leading
singular piece from g, we can write it as,
$g(t/{h^{1/\Delta}})=(t/{h^{\nu/\Delta}})f(t/h^{1/\Delta})$, where
f(x) is analytic in x.
{}From this we obtain eq. (1).

\medskip
*** The chiral restoration point is quoted in \cite{gottlieb}
to lie within an interval between two measured points.
We took the midpoint in this interval to be the critical point
and computed the $\sigma$-screening mass by linear interpolation.
For the high temperature point of this interval, both a hot
start and cold start result were given for $m_\sigma$.
We used the hot start result in Eq. (3.10).
The cold start result changes the ratio R in Eq. (3.10) to
$ R=0.844 \pm 0.040$.
Also if we use the values at the ends of the interval, we find
using the value at the lower end, $\beta =5.525$,
the ratio $R=0.813 \pm 0.043$. At the upper end, $\beta=5.55$, we find
$R=0.934 \pm 0.044$ for the hot start and $R=0.878 \pm 0.43$
for the cold start.

\bigskip

Figure Caption:

 figure 1:  Magnetization (quark condensate, $<\bar \psi \psi>$) versus
magnetic field (dimensionless quark mass, $m_q$).  The plotted points are from
lattice gauge measurements, the solid line is a fit to the leading scaling
form, and the dashed line is a fit with the first correction term
to leading scaling added.

\end{document}